\documentclass[twocolumn,prd,preprintnumbers,nofootinbib,aps,floats,floatfix,amsmath,amssymb,secnumarabic]{revtex4}
\pdfoutput=1
\usepackage[usenames,dvipsnames]{color}
\usepackage[colorlinks, citecolor=blue]{hyperref}
\usepackage{bm,graphicx}
\usepackage{psfrag}
\usepackage[normalem]{ulem}

\usepackage{epsfig,amssymb,bm, graphicx, amsmath}
\usepackage[normalem]{ulem}
\usepackage{color}

\def\be{\begin{equation}}
\def\ee{\end{equation}}
\def\ba{\begin{eqnarray}}
\def\ea{\end{eqnarray}}

\begin{document}

\title{Fossilized Gravitational Wave Relic and Primordial Clocks}
\author{Suddhasattwa Brahma}
\author{Elliot Nelson}
\author{Sarah Shandera}
\affiliation{Institute for Gravitation and the Cosmos, The Pennsylvania State University, University Park, PA 16802, USA}

\begin{abstract}
If long wavelength primordial tensor modes are coupled to short wavelength scalar modes, the scalar curvature two-point function will have an off-diagonal component. This `fossil' remnant is a signature of a mode coupling that cannot be achieved in single clock inflation. Any constraint on its presence allows a cross check of the relationship between the dynamical generation of the fluctuations and the evolution of the inflationary background. We use the example of non-Bunch Davies initial states for the tensor and scalar modes to demonstrate that physically reasonable fossils, consistent with current data, can be observable in the near future. We illustrate how the fossil off-diagonal power spectrum is a complementary probe to the squeezed limit bispectra of the scalar and tensor sectors individually. We also quantify the relation between the observable signal and the squeezed limit bispectrum for a general scalar-scalar-fossil coupling, and note the effect of superhorizon tensor modes on the anisotropy in scalar modes.
\end{abstract} 

\preprint{IGC-13/10-1}

\maketitle

\section{Introduction}
The observation of primordial gravitational waves would be an extremely important verification of the inflationary scenario and would give us a key piece of information - the energy scale of inflation. Furthermore, a measurement of a red tilt in the tensor mode power would be very strong evidence for inflation itself. However, both the characteristics of inflationary gravitational waves and the range of their observable consequences are in fact richer than just the power spectrum of B-modes in the CMB. Primordial tensor fluctuations can also affect observed Large Scale Structure statistics \cite{Jeong:2012nu, Schmidt:2012nw} and 21-cm radiation \cite{Masui:2010cz, Book:2011dz} through intrinsic distortion of the geometry as well as lensing effects. Furthermore, the tensor fluctuations sourced during inflation have self-interactions as well as interactions with the scalar fluctuations, as computed in \cite{Maldacena:2002vr, Maldacena:2011nz}. Although measuring the non-Gaussianity of just the tensor modes may be out of reach for the foreseeable future \cite{Adshead:2009bz} (or maybe not \cite{Cook:2013xea}), it has been suggested \cite{Masui:2010cz,Giddings:2011zd, Jeong:2012df,Dai:2013ikl} that in some cases the three-point correlation of two scalar modes and one tensor mode (more generally, any primordial ``fossil'' field coupled to scalar modes) is observable in Large Scale Structure or the CMB as an anisotropic contribution to the scalar power spectrum.

In single-clock inflation the squeezed limit of the tensor-scalar-scalar bispectrum, in which short-wavelength scalar modes couple to long-wavelength tensor modes, is fixed in terms of the power spectra by a consistency relation \cite{Maldacena:2002vr,Creminelli:2004yq}. The observed bispectrum vanishes in the exact squeezed limit \cite{Pajer:2013ana}, so long wavelength modes merely rescale the local background and have only a small, infrared-suppressed effect on local statistics \cite{Dai:2013kra}.

For the scalar perturbations alone, there are several ways to relax the single-clock conditions necessary for the consistency relation and to generate a physical coupling between long and short wavelength modes. All of them work by allowing additional dynamical freedom in the scalar fluctuations that is not associated purely to the evolution of the nearly de Sitter background. For example, the curvature mode can evolve outside the horizon if there are scalar fields other than the inflaton (isocurvature modes) or if a non-attractor phase of inflation preceded the usual slow-roll \cite{Namjoo:2012aa,Chen:2013aj,Chen:2013eea,Huang:2013oya,Martin:2012pe}. Allowing a non-Bunch Davies initial state for the fluctuations can also lead to a non-trivial squeezed limit for the scalar bispectrum \cite{Agullo:2011xv, Ganc:2011dy}. Depending on the range of initial states considered, the full bispectrum can also be enhanced in other configurations compared to single-field inflation where modes begin in the Bunch-Davies state \cite{Nitti:2005ym, Chen:2006nt, Holman:2007na, Das:2009sg, Meerburg:2009ys, Ashoorioon:2010xg, Agarwal:2012mq, Battefeld:2013xka, Lello:2013mfa, Sugimura:2013cra}.

There has so far not been as much theoretical effort put into trying to break the standard slow-roll relationships between the properties of the tensor fluctuations and the inflationary background, although a few proposals exist for generating a blue tensor index from inflation (solid inflation \cite{Endlich:2012pz} and `generalized G-inflation' that gives the tensors an evolving sound horizon \cite{Kobayashi:2011nu}). Here, to be illustrative, we will alter the typical inflationary signatures of  correlations beyond the power spectrum by considering a general class of non-Bunch-Davies initial states for the tensor and scalar modes. This will allow us to explore the qualitative observational features of allowing up to three separate clocks: one for the background, one for the scalar modes, and one for the tensors. We are interested in the case with a fossil signature, which is generically accompanied by a non-negligible squeezed limit bispectrum of scalars and/or tensors. Depending on how many separate clocks are physically realizable, a ``fossil'' might be accompanied by related observational signals in the scalar sector (for example, the halo bias \cite{Ganc:2012ae, Agullo:2012cs}) and/or in the gravitational sector.

The plan of the paper is as follows: In Section \ref{sec:2} we compute the fossil gravitational wave signature following the method of \cite{Jeong:2012df}. We begin with the three-point function with generalized initial states for the scalar and tensor modes and compute the signal-to-noise for the local anisotropy in the scalar power spectrum. We study its dependence on the occupation numbers for scalar and tensor excitations and on the range of excited modes. In Section \ref{sec:fossilsqueezedlimit} we consider a general scalar-scalar-fossil correlation and show the dependence of the signature on the fossil power spectrum amplitude and bispectrum amplitude and squeezed-limit, considering in particular the case of primordial tensor modes. In Section \ref{sec:quadrupole} we comment on the effect of superhorizon tensor modes on local anisotropy in the scalar power spectrum. We conclude in Section \ref{sec:conclusion}.

\section{A generalized initial state and the gravitational fossil}
\label{sec:2}
Since it is unlikely that inflation lasted forever, and since any theory of inflation is anyway likely to be only an effective description valid below some energy scale, it is quite reasonable to allow a non-Bunch-Davies initial state for primordial scalar or tensor modes. While an application of the cosmological principle suggests that it is a fine-tuning to insist that the deviation from Bunch-Davies is significant on the scales corresponding to those we observe in the CMB, it is not at all clear how to put a measure on models of inflation and how long inflation might have lasted. The possibility of deviations from Bunch-Davies is an important conceptual point about the inflationary paradigm and understanding the observational possibilities, which can be ruled out, is useful in deciding which aspects of inflation theory can be robustly tested. 

In the scalar sector, deviations from a Bunch-Davies state at the onset of inflation can result, for example, from a previous non-attractor phase \cite{Lello:2013mfa}, a previous phase with anisotropic expansion \cite{Dey:2011mj, Dey:2012qp}, tunneling from a false vacuum followed by inflation within the bubble \cite{Sugimura:2013cra,Battefeld:2013xka,Yamamoto:1996qq}, or ultraviolet completions of inflation such as loop quantum gravity \cite{Agullo:2013ai}. An effective field theory treatment of the scalar fluctuations that parametrizes non-Gaussian effects in terms of an energy scale $M$ (with $\sqrt{\epsilon}M_p\leq M>H$), should also incorporate the possibility of a generic initial state \cite{Agarwal:2012mq} for modes near the scale $M$. Less work has been put into examples of modifications of the initial state for tensor modes. Carney et. al. \cite{Carney:2011hz} found that pre slow-roll dynamics for the inflaton will not affect the quadratic action for tensor modes, which will remain in the vacuum state. However, more generically it seems reasonable to allow the initial state of both tensor and scalars to be modified in independent ways. In the next section we work out the bispectrum for one tensor mode and two scalars from the standard inflationary action but allowing non-Bunch Davies states.

\subsection{Scalar-scalar-tensor bispectrum}
\label{sec:bispectrum}

Consider an initial state for the scalar ($\zeta$) and tensor ($\gamma$) fluctuations that is a general Bogoliubov transformation of the Bunch-Davies state:
\ba
\zeta_{\mathbf{k}}(\eta)&=&\tilde{u}_k^{(s)}(\eta)a_{\mathbf{k}}+\tilde{u}_k^{(s)*}(\eta)a_{-\mathbf{k}}^\dagger, \nonumber \\
\gamma^p_{\mathbf{k}}(\eta)&=&\tilde{u}_k^{(t)}(\eta)a^p_{\mathbf{k}}+\tilde{u}_k^{(t)*}(\eta)a_{-\mathbf{k}}^{p\dagger},
\ea
where $a_{\mathbf{k}}$, $a_{\mathbf{k}}^{\dagger}$ and $a^p_{\mathbf{k}}$, $a_{\mathbf{k}}^{p,\dagger}$ are canonical creation and annihilation operators for scalar and tensor modes respectively, $p$ labels the graviton polarization, and $\tilde{u}^{(s),(t)}$ includes the Bogoliubov transformation on the scalar and tensor mode functions,
\ba\label{modefunction}
\tilde{u}^{(s)}_k(\eta)=\alpha^{(s)}_k u_k^{(s)}(\eta)+\beta^{(s)}_k u_k^{*(s)}(\eta), \nonumber \\
\tilde{u}_k^{(t)}(\eta)=\alpha^{(t)}_k u_k^{(t)}(\eta)+\beta^{(t)}_k u_k^{*(t)}(\eta).
\ea
Here, $u_k^{(s)}(\eta)=\frac{H^2}{\dot{\phi}}\frac{1}{\sqrt{2k^3}}(1+ik\eta)e^{-ik\eta}$, and $u_k^{(t)}(\eta)=\frac{H}{M_p}\frac{1}{\sqrt{k^3}}(1+ik\eta)e^{-ik\eta}$. The normalization conditions are
\ba\label{normalize}
|\alpha^{(s)}_k|^2-|\beta^{(s)}_k|^2&=&1,\nonumber\\
|\alpha^{(t)}_k|^2-|\beta^{(t)}_k|^2&=&1.
\ea
The power spectra are defined in terms of the two-point functions,
\ba
\langle\zeta_{\mathbf{k}_1}\zeta_{\mathbf{k}_2}\rangle&=&(2\pi)^3\delta^3(\mathbf{k}_1+\mathbf{k}_2)P_{\zeta}(k_1), \nonumber\\
\langle\gamma^p_{\mathbf{k}_1}\gamma^{p'}_{\mathbf{k}_2}\rangle&=&(2\pi)^3\delta^3(\mathbf{k}_1+\mathbf{k}_2)\delta^{pp'}P^p_{\gamma}(k_1).
\ea
We will define $\eta_0$ as the earliest (conformal) time where we expect the inflationary description to be valid. In specific scenarios with a pre-inflationary era, $\eta_0$ would be the beginning of standard slow-roll inflation. Alternatively, $\eta_0$ might be the earliest time when all modes of observational relevance had physical scale below some maximum energy scale $M$ we understand,  $k/a(\eta_0)\lesssim M$. The scale $M$ should be well above the Hubble scale for inflation, $M\gg H$, in order for the classical inflationary background to be valid. For modes that are well within the horizon during inflation, $|k\eta_0|\gg1$, the scalar and tensor power spectra are, respectively,
\ba
P_{\zeta}(k)&=&\frac{H^2}{2\epsilon M_p^2}\frac{1}{2k^3}|\alpha^{(s)}_k+\beta^{(s)}_k|^2
\label{Pzeta}
\ea
where $\epsilon\equiv\frac{\dot{\phi}^2}{2H^2 M_p^2}$, and 
\be
P^p_{\gamma}(k)=\frac{4H^2}{M_p^2}\frac{1}{2k^3}|\alpha^{(t)}_k+\beta^{(t)}_k|^2\;.
\label{Ptensor}
\ee
In both cases, the $\beta^{(s),(t)}(k)\rightarrow0$ sufficiently rapidly for $k>a(\eta_0)M$ to avoid any back reaction that would spoil the inflationary background. The tensor-to-scalar ratio is
\be\label{TSratio}
r\equiv\frac{\sum_p P^p_{\gamma}}{P_{\zeta}}=16\epsilon\frac{|\alpha^{(t)}_k+\beta^{(t)}_k|^2}{|\alpha^{(s)}_k+\beta^{(s)}_k|^2}.
\ee
The extra factors must be consistent with the observed near scale-invariance of the scalar power spectrum but can alter the standard slow-roll consistency relations. In particular, some functions $\alpha^{(s),(t)}$, $\beta^{(s),(t)}$ could suppress the tensor-to-scalar ratio enough to allow high energy scale or large field models of inflation to be consistent with {\it Planck} satellite constraints \cite{Ashoorioon:2013eia}. Given the normalization condition Eq. \eqref{normalize} we can write
\ba
|\alpha^{(s)}_k+\beta^{(s)}_k|^2&=&1+2|\beta^{(s)}_k|^2\\
&&+2\sqrt{|\beta^{(s)}_k|^2(|\beta^{(s)}_k|^2+1)}\cos\Theta^{(s)}, \nonumber \\
|\alpha^{(t)}_k+\beta^{(t)}_k|^2&=&1+2|\beta^{(t)}_k|^2\nonumber\\
&&+2\sqrt{|\beta^{(t)}_k|^2(|\beta^{(t)}_k|^2+1)}\cos\Theta^{(t)}, \nonumber
\ea
where $\Theta^{(s)}$ is the relative phase between $\alpha^{(s)}_k$ and $\beta^{(s)}_k$, and $\Theta^{(t)}$ is defined similarly.

During inflation the tensor ($\gamma^p$ with two polarizations $p$) and scalar ($\zeta$) fluctuations are coupled gravitationally, giving rise to a three point correlation
\be
\langle\gamma^p(\mathbf{k}_1)\zeta(\mathbf{k}_2)\zeta(\mathbf{k}_3)\rangle\equiv(2\pi)^3\delta^3(\sum \mathbf{k}_i)B_p(k_1,k_2,k_3). \nonumber
\ee
Following Maldacena's calculation \cite{Maldacena:2002vr} using the in-in formalism \cite{Weinberg:2005vy} for the Bunch-Davies case, and using the modified mode functions in Eq. \eqref{modefunction}, we find that for modified initial states
\begin{widetext}
\ba
B_p(k_1,k_2,k_3)&=&
4\frac{H^6}{M_p^2\dot{\phi}^2}\frac{1}{\prod 2k_i^3}\epsilon^p_{ij}k_2^ik_3^j \int_{\eta_0}^0 -i\frac{d\eta}{\eta^2}\;(\alpha^{(t)}_{k_1}+\beta^{(t)}_{k_1})(\alpha^{(t)*}_{k_1}(1-ik_1\eta)e^{ik_1\eta}+\beta^{(t)*}_{k_1}(1+ik_1\eta)e^{-ik_1\eta})\nonumber\\
&&\times(\alpha^{(s)}_{k_2}+\beta^{(s)}_{k_2})(\alpha^{(s)*}_{k_2}(1-ik_2\eta)e^{ik_2\eta}+\beta^{(s)*}_{k_2}(1+ik_2\eta)e^{-ik_2\eta})\nonumber\\
&&\times(\alpha^{(s)}_{k_3}+\beta^{(s)}_{k_3})(\alpha^{(s)*}_{k_3}(1-ik_3\eta)e^{ik_3\eta}+\beta^{(s)*}_{k_2}(1+ik_3\eta)e^{-ik_3\eta})
\nonumber\\&&+\text{c.c.}\nonumber\\
&=& 4\frac{H^6}{M_p^2\dot{\phi}^2}\frac{1}{\prod 2k_i^3}\epsilon^p_{ij}k_2^ik_3^j (\alpha^{(t)}_{k_1}+\beta^{(t)}_{k_1})(\alpha^{(s)}_{k_2}+\beta^{(s)}_{k_2})(\alpha^{(s)}_{k_3}+\beta^{(s)}_{k_3})\nonumber\\
&&\times\bigg[
\alpha^{(t)*}_{k_1}\alpha^{(s)*}_{k_2}\alpha^{(s)*}_{k_3}I(k_1,k_2,k_3;\eta_0)+\beta^{(t)*}_{k_1}\alpha^{(s)*}_{k_2}\alpha^{(s)*}_{k_3}I(-k_1,k_2,k_3;\eta_0)\nonumber\\
&&+\alpha^{(t)*}_{k_1}\beta^{(s)*}_{k_2}\alpha^{(s)*}_{k_3}I(k_1,-k_2,k_3;\eta_0)+\alpha^{(t)*}_{k_1}\alpha^{(s)*}_{k_2}\beta^{(s)*}_{k_3}I(k_1,k_2,-k_3;\eta_0)\nonumber\\
&&+\beta^{(t)*}_{k_1}\beta^{(s)*}_{k_2}\alpha^{(s)*}_{k_3}I(-k_1,-k_2,k_3;\eta_0)+\beta^{(t)*}_{k_1}\alpha^{(s)*}_{k_2}\beta^{(s)*}_{k_3}I(-k_1,k_2,-k_3;\eta_0)\nonumber\\
&&+\alpha^{(t)*}_{k_1}\beta^{(s)*}_{k_2}\beta^{(s)*}_{k_3}I(k_1,-k_2,-k_3;\eta_0)+\beta^{(t)*}_{k_1}\beta^{(s)*}_{k_2}\beta^{(s)*}_{k_3}I(-k_1,-k_2,-k_3;\eta_0)\bigg]\nonumber\\&&+\text{c.c.}, \ \ \ \  \label{bispectrum}
\ea

where

\ba\label{I}
I(p_1,p_2,p_3)&=&-(p_1+p_2+p_3)+\frac{\sum_{i>j}p_i p_j}{p_1+p_2+p_3}+\frac{p_1p_2p_3}{(p_1+p_2+p_3)^2}\nonumber\\
&&-e^{i(p_1+p_2+p_3)\eta_0}\Big[\frac{\sum_{i>j}p_i p_j}{p_1+p_2+p_3}+\frac{p_1p_2p_3(1-i(p_1+p_2+p_3)\eta_0)}{(p_1+p_2+p_3)^2}+\frac{i}{\eta_0}\Big], \ \ \ \ \ \ 
\ea
and $p_t=p_1+p_2+p_3$. For the permutations in Eq. \eqref{bispectrum}, $p_t$ will be either $k_t$ or combinations such as $k_1-k_2+k_3$. 
\end{widetext}

To evaluate the time integrals in the first expression of Eq.(\ref{bispectrum}), we have used, for example, that the integral for the $\alpha\alpha\alpha$ permutation term is

\ba
\mathcal{I}_{\alpha\alpha\alpha}&\equiv&-\int_{\eta_0}^0 i\frac{d\eta}{\eta^2}\prod_i (1-ik_i\eta)e^{ik_i\eta}\nonumber\\
&=&e^{ik_t\eta}\left\lbrace \frac{\sum_{i<j} k_ik_j}{k_t}+\frac{k_1k_2k_3}{k_t^2}(1-i k_t \eta)+\frac{i}{\eta} \right\rbrace \bigg|_{\eta=\eta_0}^{0},\nonumber\\
&=&I(k_1,k_2,k_3)+\lim_{\eta\rightarrow0}\frac{i}{\eta}e^{ik_t\eta}.\nonumber
\ea
The divergent $\frac{i}{\eta}$ piece comes from the $\frac{1-ik_t\eta}{\eta^2}$ part of the integrand and is purely imaginary and independent of the $k_i$ in the $\eta\rightarrow0$ limit. We can therefore factor this piece outside of the permutations, with a change of sign in each of the conjugate pieces to see that the Bogoliubov coefficients of this imaginary, divergent, piece multiply to a real quantity, 
\ba
&&(\alpha^{(t)}_{k_1}+\beta^{(t)}_{k_1})(\alpha^{(s)}_{k_2}+\beta^{(s)}_{k_2})(\alpha^{(s)}_{k_3}+\beta^{(s)}_{k_3})
\nonumber\\
 &&\times\,[(\alpha^{(t)}_{k_1}+\beta^{(t)}_{k_1})(\alpha^{(s)}_{k_2}+\beta^{(s)}_{k_2})(\alpha^{(s)}_{k_3}+\beta^{(s)}_{k_3})]^*\;,
\ea
which is canceled by its complex conjugate in the first expression of Eq.(\ref{bispectrum}), leading to the second expression.

For Bunch-Davies initial states, $\alpha^{(s)}_k=\alpha^{(t)}_k=1$ and $\beta^{(s)}_k=\beta^{(t)}_k=0$, Eq. \eqref{bispectrum} reduces to Eq. $(4.10)$ of \cite{Maldacena:2002vr} if we take $\eta_0\rightarrow-\infty$ with the contour prescription used there.\footnote{We have an extra factor of two; this comes from the tensor power spectrum Eq. \eqref{Ptensor}, which is consistent with the consistency relation $r=16\epsilon$ in the Bunch-Davies case and differs from that in \cite{Maldacena:2002vr} by a factor of two.}

It follows from Eq. \eqref{bispectrum} that an excited initial state can lead to a large three-point correlation in the squeezed limit in the coordinate frame appropriate for late-time observers.
It was shown in \cite{Pajer:2013ana} that moving to this frame, described with conformal Fermi Normal Coordinates, introduces another term in the three-point function in the squeezed limit $k_L\equiv k_1\ll k_2\simeq k_3\equiv k_S$,
\ba
B_p^{\text{obs}}(k_L,k_2,k_3)&=&B_p(k_L,k_2,k_3)\\
&+&\frac{1}{2}P^p_{\gamma}(k_L)P_{\zeta}(k_S)\epsilon_{ij}^p k_S^i k_S^j \frac{\partial\ln P_{\zeta}(k_S)}{\partial\ln k_S}. \nonumber
\label{Bobs}
\ea
When the consistency relation of single-clock inflation is satisfied the additional term in Eq. \eqref{Bobs} cancels the first term in the exact squeezed limit (to order $(k_L/k_s)^2$). In that case there is no physical correlation with long wavelength modes and the apparently nonvanishing squeezed limit is a gauge artifact (see \cite{Tanaka:2011aj} for an earlier version of this argument). With a general initial state, the consistency relation need not hold: for a finite range of modes, a small physical coupling between long and short wavelength modes is consistent with inflationary expansion of the background even though the correlation is not {\it generated} by the background\footnote{In other words, the consistency relation in the \textit{exact} squeezed limit $k_L/k_S\rightarrow0$ is not violated. We are only computing correlation functions for modes that are subhorizon at the onset of the inflationary era, $|k_L\eta_0|\gtrsim1$, and longer wavelength modes are treated as part of the background. Taking $\eta_0$ farther into the past tightens the back-reaction bound on $\beta^{(s)}$ \cite{Holman:2007na}, so that taking $\eta_0\rightarrow-\infty$ returns us to the Bunch-Davies state, $\beta^{(s)}=0$. Sufficiently short wavelength, observable modes will also have $\beta^{(s)}=0$.}. The permutations in Eq. \eqref{bispectrum} proportional to $\beta^{(i)}$ can dominate the squeezed limit on some scales, yielding a physical coupling between a long wavelength tensor mode with the short wavelength fluctuations that can be seen in the observer's reference frame. The details of the shape, however, depend on the $k-$dependence of the $\alpha$ and $\beta$ coefficients. In the $\beta_k^{(s)}=\beta_k^{(t)}=0$ case these terms (and hence the observed squeezed limit) vanish \cite{Maldacena:2002vr,Pajer:2013ana}.

\subsection{Enhanced fossil signature in locally anisotropic scalar fluctuations}
\label{sec:estimator}

The scalar-scalar-tensor correlation can be used to estimate the magnitude of an unseen tensor mode $\gamma^p(\mathbf{K})$ through observations of local anisotropy in scalar modes. In general, if scalar curvature perturbations $\zeta$ are coupled to an unobservable fossil field with an isotropic scalar-scalar-fossil three-point function, the local power spectrum evaluated in the presence of a tensor field realization is \cite{Jeong:2012df,Dai:2013kra}
\ba
\label{offdiagonal}
\langle\zeta(\mathbf{k}_1)\zeta(\mathbf{k}_2)\rangle_{\gamma}&=&(2\pi)^3\delta^3(\mathbf{k}_1\mathbf{k}_2)P_{\zeta}(k)\nonumber\\
&+&\sum_p \int\frac{d^3K}{(2\pi)^3}(2\pi)^3\delta^3(\mathbf{k}_1+\mathbf{k}_2+\mathbf{K})\nonumber\\
&&\times\;\gamma_p^{*}(\mathbf{K})\frac{B_p(K,k_1,k_2)}{P^p_{\gamma}(K)}.
\ea 

Each pair of scalar modes whose momenta add to $\mathbf{K}$ can be used to give an estimator for $\gamma^p(\mathbf{K})$. The minimum variance estimator obtained from all such pairs has some uncertainty, which is quantified by the noise power spectrum $P_p^n(K)$, as defined in Eq. $(5)$ of \cite{Jeong:2012df}:
\ba
[P_p^n(K)]^{-1}=\sum_{\mathbf{k}} \dfrac{B_p^2(K,k,|\mathbf{K}-\mathbf{k}|)}{2VP^p_{\gamma}(K)^2P^{\text{tot}}_{\zeta}(k)P^{\text{tot}}_{\zeta}(|\mathbf{K}-\mathbf{k}|)}, \label{noisepowerdef}
\ea
where $P_{\zeta}^{\text{tot}}(k)$ is the measured scalar power spectrum, including signal and noise, and $V\equiv(2\pi/k_{\text{min}})^3$ is the volume of the survey. Going to the continuous limit $\sum_{\mathbf{k}}\rightarrow V\int d^3k/(2\pi)^3$ and making use of Eqs. \eqref{Pzeta}, \eqref{Ptensor} and \eqref{bispectrum}, we find
\ba
[P_p^n(K)]^{-1}&=&|\alpha^{(t)}_K+\beta^{(t)}_K|^{-4}\int_{k_{\text{min}}}^{k_{\text{max}}} \frac{d^3 k}{(2\pi)^3}|\alpha^{(s)}_k+\beta^{(s)}_k|^{-2}\nonumber\\
&&\times|\alpha^{(s)}_{|\mathbf{K}-\mathbf{k}|}+\beta^{(s)}_{|\mathbf{K}-\mathbf{k}|}|^{-2}
\dfrac{k^4 \sin^4\theta \cos^2 2\phi}{8k^3|\mathbf{K}-\mathbf{k}|^3}\nonumber\\
&\times&\left[(\alpha^{(t)}_{K}+\beta^{(t)}_{K})(\alpha^{(s)}_{k}+\beta^{(s)}_{k})\right.\nonumber\\
&&\times(\alpha^{(s)}_{|\mathbf{K}-\mathbf{k}|}+\beta^{(s)}_{|\mathbf{K}-\mathbf{k}|})\tilde{I}\left.+\text{c.c.}\right]^2, \label{noisepower}
\ea
where $\tilde{I}$ denotes the quantity in brackets in Eq. \eqref{bispectrum}. Here we have followed \cite{Jeong:2012df} by approximating $P_{\zeta}(k)/P_{\zeta}^{\text{tot}}(k)\simeq1$ for $k<k_{\text{max}}$ and $P_{\zeta}(k)/P_{\zeta}^{\text{tot}}(k)\simeq0$ for $k>k_{\text{max}}$, where $k_{\text{max}}$ is the smallest scale at which the power spectrum can be measured.  We have also taken $\mathbf{K}$ to be in the $\hat{\mathbf{z}}$ direction, so that the polarization tensor $\epsilon^p_{ij}$ takes the form $\epsilon^{+}_{xx}=-\epsilon^{+}_{yy}=1$ for the $+$ polarization, and $\epsilon^{\times}_{xy}=\epsilon^{\times}_{yx}=1$ for the $\times$ polarization, with all other components zero. This yields the factor $\cos^2 2\phi$ for the $+$ polarization, as shown above; the $\times$ polarization yields a factor of $\sin^2 2\phi$ instead, but the integral is identical.

We will take $\alpha^{(s),(t)}_k$ and $\beta^{(s),(t)}_k$ to be constant for the observable range $k_{\text{min}}<k<k_{\text{max}}$, and consider possible $k$-dependence later in this section.\footnote{In the limit of large $k$, $\beta^{(s),(t)}_k$ must approach zero sufficiently quickly or vanish above some scale so that back-reaction constraints on the energy density are met \cite{Holman:2007na,Flauger:2013hra}.} We will also take the time $\eta_0$ at which we specify the initial state for inflation to be early enough so that all modes of interest are well inside the horizon, $|k_i\eta_0|\gg1$. Furthermore, due to the limited resolution of modes in a finite volume, only configurations for which $k_1-k_2+k_3\gtrsim k_{\text{min}}$ (and similarly for other permutations) will contribute to the observable signal. Since we assume $|k_{\text{min}}\eta_0|\gg1$, the terms in the second line of Eq. \eqref{I} oscillate rapidly and will average to zero when we integrate over $k$, so we discard them from now on. The $k\rightarrow k_{\text{max}}$ limit dominates the integral\footnote{There is also a divergence in the collinear limit for $K<k$, but upon cutting off the integral with $k_{\text{min}}$, this gives a subdominant contribution, which is completely negligible in the $K\rightarrow k_{\text{min}}$ limit that dominates the final result, Eq. \eqref{sigmat}.}, and the second and third terms\footnote{In fact, only the third term will matter, but we will include both for completeness.} of Eq. \eqref{I} are enhanced by a factor of $\frac{k}{K}$ in the middle four permutations in Eq. \eqref{bispectrum}, so the bispectrum scales as $k^{-2}K^{-4}$ in the squeezed limit $K\ll k$. As long as $\beta_k$ is not too small (which we will quantify below), these terms dominate the integral. (The coordinate transformation term from Eq. \eqref{Bobs} scales as $k^{-3}K^{-3}$, as do the subdominant permutations in Eq. \eqref{bispectrum}, and can thus be ignored.)

The angular dependence in the denominator of these terms, eg. $k_1-k_2+k_3=K(1+\cos\theta)$, contributes a divergence in the collinear or flattened limit. Dropping factors of (constant) $\alpha$ and $\beta$ for now and cutting off the angular integral at $\theta_{\text{min}}\simeq k_{\text{min}}/K$ and $\theta_{\text{max}}\simeq\pi-k_{\text{min}}/K$, we find
\ba
[P_p^n(K)]^{-1}&\propto&\frac{\pi}{40(2\pi)^3}\frac{k_{\text{max}}^5}{K^2}\int_{\theta_{\text{min}}}^{\theta_{\text{max}}} d\theta \sin^5\theta\left(\frac{4}{\sin^4\theta}\right)^2\nonumber\\
&\propto&\frac{16\pi}{40(2\pi)^3}\frac{k_{\text{max}}^5}{K^2}\left(\frac{K}{k_{\text{min}}}\right)^2
\label{Pnpintegral}
\ea
Restoring factors of $\alpha$ and $\beta$ and using the normalization conditions, Eq. \eqref{normalize}, we have
\be
[P_p^n(K)]^{-1}=\frac{k_{\text{max}}^3}{20\pi^2}\left(\frac{k_{\text{max}}}{k_{\text{min}}}\right)^2F^{-1}(\beta^{(s)},\beta^{(t)},\Theta^{(s)},\Theta^{(t)})\label{PnpFinal},
\ee
where
\ba
&&F^{-1}(\beta^{(s)},\beta^{(t)},\Theta^{(s)},\Theta^{(t)})\nonumber\\
&&\hspace{3mm}=\beta^{(s)2}(1+\beta^{(s)2}) \nonumber\\
&&\hspace{5mm} \times \left[\Big(1+2\beta^{(t)2}+2\sqrt{\beta^{(t)2} (\beta^{(t)2}+1)}\cos\Theta^{(t)}\Big) \right.\nonumber\\
&&\left.\hspace{9mm}\times \Big(1+2\beta^{(s)2}+2\sqrt{\beta^{(s)2}(\beta^{(s)2}+1)}\cos\Theta^{(s)}\Big)\right]^{-2}\nonumber\\
&&\hspace{5mm}\times\left[\Big(\sqrt{1+\beta^{(s)2}}e^{i\Theta^{(s)}}+\beta^{(s)}\Big)^2\right.\nonumber\\
&&\hspace{9mm}\times\Big(\sqrt{1+\beta^{(t)2}}e^{i\Theta^{(t)}}+\beta^{(t)}\Big)\nonumber\\
&&\left.\hspace{9mm}\times
\Big(\sqrt{1+\beta^{(t)2}}e^{-i\Theta^{(t)}}-\beta^{(t)}\Big)e^{-i\Theta^{(s)}}\text{ + c.c.}\right]^2, \nonumber\\
\ea
where $\beta^{(s)}\equiv|\beta_k^{(s)}|$, $\beta^{(t)}\equiv|\beta^{(t)}_k|$, and $\Theta^{(s)}$ and $\Theta^{(t)}$ are the relative phases between $\alpha^{(s)}$ and $\beta^{(s)}$, and $\alpha^{(t)}$ and $\beta^{(t)}$, respectively. Note that the factor of $({k_{\text{max}}/k_{\text{min}}})^2$ in Eq. \eqref{PnpFinal} comes from the $k^{-2}K^{-4}$ scaling of the non-Bunch-Davies bispectrum as described above.

The estimates for each mode $\gamma^p(\mathbf{K})$ can be combined to give a minimum variance estimator for the power $A_{\gamma}\equiv K^3 P^p_{\gamma}(K)=2\frac{H^2}{M_p^2}|\alpha^{(t)}_k+\beta^{(t)}_k|^2$ in tensor fluctuations (we assume a nearly scale invariant tensor power spectrum). The overall variance is obtained by summing over the inverse variances from each $K$ \cite{Jeong:2012df}:
\be
\sigma_{\gamma}^{-2}\equiv\frac{1}{2}\sum_{\mathbf{K},p}\left[K^3 P_p^n(K)\right]^{-2}.
\label{sigmat}
\ee
We see from Eq. \eqref{PnpFinal} that $P_p^n(K)$ is independent of $K$ in the regime we are considering, and the sum over polarizations simply gives a factor of two, so the variance of the estimator for $A_{\gamma}$ is given by $\sigma_{\gamma}^2=\frac{3}{4\pi}k_{\text{min}}^6(P_p^n)^2$. If we require a $3\sigma$ detection of $A_{\gamma}$, the minimum detectable amplitude for the tensor power spectrum is
\be
3\sigma_{\gamma}=30\pi\sqrt{3\pi}\left(\frac{k_{\text{max}}}{k_{\text{min}}}\right)^{-5}F(\beta^{(s)},\beta^{(t)},\Theta^{(s)},\Theta^{(t)}).   \label{threshold}
\ee
The minimum value of $k_{\text{max}}/k_{\text{min}}$ needed for a survey capable of detecting primordial tensor modes of a given amplitude is found by setting the detection threshold of Eq. \eqref{threshold} equal to the tensor amplitude $A_{\gamma}$.

\begin{figure*}

\centering

\includegraphics[width=.9\textwidth]{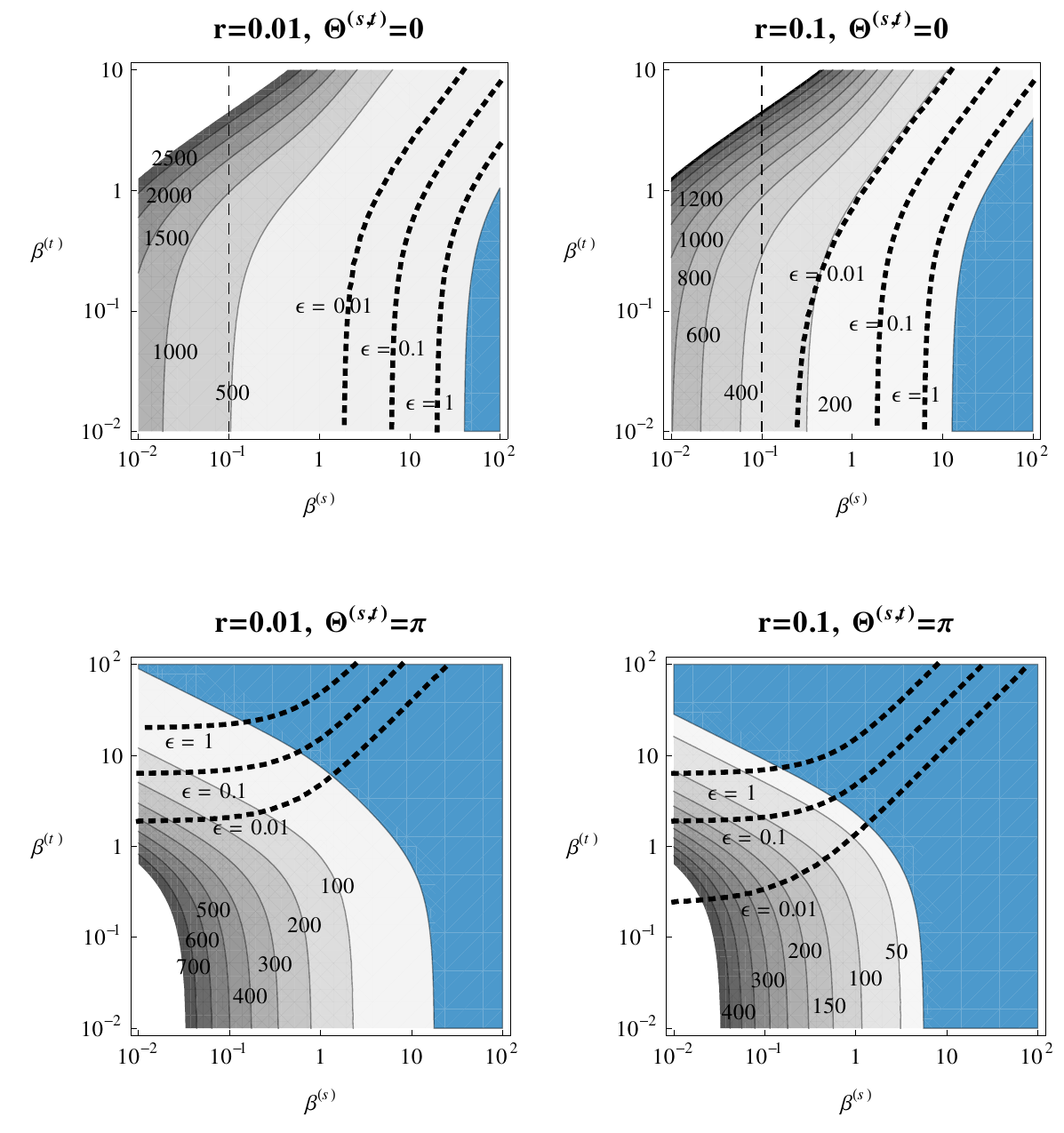}
\caption{Contour plot of the minimum survey size $k_{\text{max}}/k_{\text{min}}$ needed to detect the effect of primordial tensor fluctuations through off-diagonal contributions to the scalar power spectrum, in terms of scalar and tensor Bogoliubov parameters $\beta^{(s)},\beta^{(t)}$, for different values of the tensor-to-scalar ratio $r$ and non-Bunch-Davies phases $\Theta^{(s)},\Theta^{(t)}$. The scales $k_{\text{min}}$ and $k_{\text{max}}$ are the longest and shortest observable scales (at which scalar modes are excited). The dashed line at $\beta^{(s)}=0.1$ indicates the back-reaction bound $\beta^{(s)}\lesssim0.1$ (this does not apply in the $\Theta^{(s)}=\pi$ case \cite{Flauger:2013hra}). The dark shaded region is ruled out by the Planck constraint $f^{\rm NBD2}_{\rm NL}=0.2\pm0.4$. For $\Theta^{(s),(t)}=\pi$ the power spectra decrease rather than increase with $\beta^{(s),(t)}$, leading to different behavior.\label{phases00}}
\end{figure*}

In ~\ref{phases00} we show the dependence of the survey size required to detect tensor modes on the initial state parameters $\beta^{(s)}$ and $\beta^{(t)}$, for two values of the tensor-to-scalar ratio $r$ and two pairs of values for the angles $\Theta^{(s)},\Theta^{(t)}$. For a given survey volume $V=(2\pi/k_{\text{min}})^3$, the contour lines determine values of $k_{\text{max}}$ required to make a detection. (The effect of the overall bispectrum and tensor power spectrum amplitudes on the signal is somewhat obscured by their complicated dependence on the Bogoliubov transformation parameters, and is shown more simply in Figure \ref{PvsB} below.)

As expected, the three-point correlation is stronger for larger $\beta^{(s)}$, resulting in a smaller minimum survey size. Although one might expect the increase in the tensor power spectrum $P_{\gamma}$ for larger $\beta^{(t)}$ to universally decrease the minimum survey size, a larger $\beta^{(t)}$ also affects the variance of the estimator for $P_{\gamma}$ and can, depending on the phases, increase or decrease the minimum survey size.

In order to have inflation we require $\epsilon<1$, which constrains the $(\beta^{(s)},\beta^{(t)})$ parameter space if we also fix $r$ in Eq. \eqref{TSratio}. This is shown in Figure~\ref{phases00} in the form of curves of fixed $\epsilon$. Furthermore, it was shown in \cite{Flauger:2013hra} that requiring (i) that the energy density from the excited modes not spoil the slow roll evolution, and (ii) that the observed power spectrum be nearly scale-invariant, rules out large occupation numbers $\beta^{(s)}\gg1$ but allows for $\beta^{(s)}\lesssim0.1$. This bound is indicated in Figure \ref{phases00} as a dashed line, and is relaxed in the $\Theta^{(s)}=\pi$ case \cite{Flauger:2013hra}.

Planck constraints on non-Gaussianity \cite{Ade:2013ydc} impose an observational bound on $\beta^{(s)}$ by constraining $f_{\rm NL}^{\rm NBD2}=0.2\pm0.4$, where $f_{\rm NL}^{\rm NBD2}$ parameterizes the amplitude of the bispectrum.\footnote{Note that in Eq. (14) of \cite{Ade:2013ydc}, $f_{\rm NL}^{\rm NBD2}$ is written as $f_{\rm NL}^{\rm NBD1}$.} This parameter is related to the squeezed limit of the bispectrum,\footnote{In \cite{Ade:2013ydc}, $f_{\rm NL}^{\rm NBD2}$ is the amplitude for a slightly different bispectrum template than that obtained from a Bogoliubov initial state, which has additional dependence on $\beta^{(s)}$ and $\Theta^{(s)}$ that mildly affects the momentum dependence, although the behavior in the squeezed and flattened limits is the same. The constraints shown in ~\ref{phases00} are therefore approximate.}
\ba\label{fNLdef}
f_{\rm NL}^{\rm}\hspace{-.15cm}&\equiv&\hspace{-.25cm}\lim_{k_L\ll k_S}\frac{1}{4}\frac{B_{\zeta}(k_L,|\mathbf{k}_S-\mathbf{k}_L/2|,|-\mathbf{k}_S-\mathbf{k}_L/2|)}{P_{\zeta}(k_L)P_{\zeta}(k_S)},\hspace{.5cm}
\ea
by a factor of $f_{\rm NL}^{\rm NBD2}=\mathcal{O}(1)\frac{k_L}{k_S}f_{\rm NL}$. In Eq. \eqref{fNLdef}, $B_{\zeta}$ is the $\langle\zeta^3\rangle$ bispectrum \cite{Agullo:2012cs,Ganc:2011dy} for the initial state in Eq. \eqref{modefunction}. Taking the squeezed limit we find\footnote{We omit $\mathcal{O}(1)$ factors from the angular dependence in the bispectrum.}
\ba
f_{\rm NL}^{\rm}\hspace{-.25cm}&=&\hspace{-.15cm}
\left\{
  \begin{array}{ll}
2\epsilon|\beta^{(s)}|\frac{k_S}{k_L}\cos{\Theta^{(s)}}, & \hspace{0mm} |\beta^{(s)}|\ll1 \\
\frac{1}{2}\epsilon\frac{k_S}{k_L}f(\Theta^{(s)}), & \hspace{0mm} |\beta^{(s)}|\gg1, \ \Theta^{(s)}\neq\pi  \\
-8\epsilon|\beta^{(s)}|^4\frac{k_S}{k_L}, & \hspace{0mm} |\beta^{(s)}|\gg1, \ \Theta^{(s)}=\pi,
  \end{array}
 \right.\hspace{.5cm}
 \label{fNLNBD}
\ea
where $f(\Theta^{(s)})=(3+2\cos\Theta^{(s)}-\cos 2\Theta^{(s)})/(1+\cos\Theta^{(s)})^2$. From Eq. \eqref{TSratio} we see that for a given $r$ the Planck constraints on $f_{\rm NL}^{\rm NBD2}$ impose a bound on the Bogoliubov parameters. This bound is also shown in Figure \ref{phases00} (which includes the full squeezed-limit dependence on $\beta^{(s)}$, although we have only shown the limiting cases in Eq. \eqref{fNLNBD}).

Note that depending on the phases, the dependence on $\beta^{(s)}$ and $\beta^{(t)}$ can be quite different \cite{Flauger:2013hra}. For $\Theta^{(s)}=0$ the theoretical constraint from requiring $\epsilon$ to be small is stronger than the observational constraint from Planck bounds on non-Gaussianity, consistent with the conclusions of \cite{Aravind:2013lra,Flauger:2013hra} that observable non-Gaussianity is not expected in the CMB from Bogoliubov initial states. On the other hand, for $\Theta^{(s)}=\pi$ the observational constraints play a significant role. 

In the case of a smaller speed of sound $c_s<1$, the $\langle\zeta^3\rangle$ bispectrum receives a new contribution that is zeroth order in slow roll \cite{Agarwal:2012mq}, with amplitude $f_{\rm NL}^{\rm}\propto\frac{1-c_s^2}{c_s^2}\frac{k_S}{k_L}$. This results in a stronger constraint on $\beta^{(s)}$ for smaller $c_s$. For $|\beta^{(s)}|\ll1$ the amplitude is proportional to $|\beta^{(s)}|$, so for $c_s=0.02$, saturating the constraint from Planck on equilateral and orthogonal non-Gaussianity \cite{Ade:2013ydc}, we have $|\beta^{(s)}|\lesssim10^{-2}$.

Returning to the present calculation, note that for \textit{very} small $\beta^{(s)}$, all terms with $\beta^{(s)}$ in the bispectrum, Eq. \eqref{bispectrum}, become negligible, and the threshold amplitude appears to be $3\sigma_{\gamma}=30\pi\sqrt{3\pi}(k_{\text{max}}/k_{\text{min}})^{-3}$ \cite{Jeong:2012df}. In this limit the coordinate transformation Eq. \eqref{Bobs} to the observer's frame of reference becomes significant and the squeezed-limit signal vanishes. In order for the non-Bunch-Davies terms to dominate in Eq. \eqref{noisepower} we need $F^{-1}\propto\beta^{(s)2}$ for $\beta^{(s)}\ll1$, so
\be
(\beta^{(s)})^2\left(\frac{k_{\text{max}}}{k_{\text{min}}}\right)^2\gg1. \label{betamin}
\ee
If we were to include the other terms in Eq. \eqref{noisepower} in the above calculation, we would have found fractional corrections to the result Eq. \eqref{threshold} of order $\frac{1}{\beta^{(s)}}\frac{k_{\text{min}}}{k_{\text{max}}}$.

In the case of a sharp cutoff $\Lambda<k_{\text{max}}$ within the observable range, above which scalar modes are not excited, all terms with $\beta^{(s)}_k$ vanish for $k>\Lambda$, so we cut off the integral in Eq. \eqref{noisepower} at $\Lambda$.  The final result is then modified to $\sigma_{\gamma}\propto(\Lambda/k_{\text{min}})^{-5}$, so the plots in Figure \ref{phases00} indicate a lower bound on the required $k_{\text{max}}$, which is saturated for $\beta_k^{(s)}$ nonzero over the entire range of observed modes. Alternatively, the contour lines can be thought of as showing $\Lambda/k_{\text{min}}$, giving the scale $\Lambda$ up to which scalar modes in the initial state would have to be excited to leave a detectable signature,\footnote{We assume tensor modes are also excited up to this scale; for tensor modes excited over fewer scales, the dependence on the occupation number $\beta^{(t)}$ would be weakened.} assuming it is possible to probe this scale observationally. The signal from the non-Bunch-Davies modes is dominant as long as the terms considered above in $P_p^n(\mathbf{K})$ still dominate the integral. This is true if
\be
(\beta^{(s)})^2\frac{\Lambda^5}{k_{\text{max}}^3k_{\text{min}}^2}\gg1. \label{betamin2}
\ee
In comparison to Eq. \eqref{betamin}, an extra factor of $(\Lambda/k_{\text{max}})^5$ comes from the change in the upper limit. If $\Lambda$ is too small this condition will no longer be satisfied: there will be too few excited modes in the observable range to produce a detectable signal.

We can also consider a power law parameterization, $|\beta_k|^2=\beta^2(k/k_*)^{-\delta}$, where $k_*$ functions like the cutoff $\Lambda$. If we consider the limit of small $\beta$, so $\mathcal{O}(\beta^2)$ terms in the bispectrum can be dropped, then we simply pick up an extra factor of $(k/k_*)^{-\delta}$ in the integral in Eq. \eqref{noisepower}, which leads to an extra factor of $(k_*/k_{\text{max}})^{-\delta}$ in Eq. \eqref{threshold}, and a factor $(k_{\text{max}}/k_*)^{-\delta}$ in Eq. \eqref{betamin}. The step function therefore mimics the power law behavior for $\delta=5$.

\section{General condition for scalar-scalar-fossil correlation}
\label{sec:fossilsqueezedlimit}

Although we have computed the very specific effect of a modified initial state with Bogoliubov coefficients, we can easily generalize the result to see the effect of long-wavelength fossil modes on local scalar fluctuations more generally. We will consider the squeezed limit of a factorizable three-point function with power law dependence, and compute $\sigma_f$ (in the following we will change `$\gamma$' subscripts to `$f$' for a generic fossil field). The squeezed limit is parameterized by an amplitude $f_{\zeta\zeta f}$, with additional power law dependence for the long and short-wavelength modes,
\be\label{squeezedbispectrum}
B(k_L,k_2,k_3)=f_{\zeta\zeta f}P_f(k_L)P_{\zeta}(k_S)\left(\frac{k_S}{k_p}\right)^{m_S}\left(\frac{k_L}{k_p}\right)^{m_L},
\ee
where $k_L\ll k_2\simeq k_3\equiv k_S$, and we ignore any angular dependence. For a nearly scale-invariant power spectrum and bispectrum, $n_f\simeq0$ and $m_S+m_L\simeq0$. We parameterize the power spectrum as $P_f(k)\equiv\frac{A_f}{k^3}\left(\frac{k}{k_p}\right)^{n_f-3}$. If we require a detection of significance of $\alpha\equiv A_f/\sigma_f$ standard deviations, then we arrive at an inequality relating the survey size, strength of the fossil correlation, amplitude of curvature and fossil fluctuations, and runnings of the tensor power spectrum and squeezed-limit three-point function,
\be
\frac{1}{C}f_{\zeta\zeta f}^2 A_f\left(\frac{k_{\text{max}}}{k_{\text{min}}}\right)^{2m_S+3}>\alpha
 \label{squeezedinequality}
\ee
where
\be
C\equiv 2\pi^{3/2}(2m_S+3)(3-4m_L-2n_f)^{1/2},
\ee
and we have chosen $k_p=k_{\text{min}}$. We have assumed that the integral over $k$ for the noise power spectrum is dominated by the upper limit, and the integral over $K$ for $\sigma_f^{-2}$ is dominated by the lower limit, so that the signal comes from the squeezed limit $K\ll k$; this requires that $m_S>-3/2$ and $4m_L+2n_f-3<0$.

Eq. \eqref{squeezedinequality} shows that a detectable fossil signal would require either a nonzero positive value
of $m_S$ (and thus for scale invariant bispectra a strong divergence in the $K\rightarrow0$ limit) as in the case above, a large bispectrum amplitude, and a sufficiently large amplitude of fossil field fluctuations. If the bispectrum is not proportional to $A_f$, that is, if $f_{\zeta\zeta f}$ as defined above depends on $A_f$, the effect of the fossil amplitude on the signal strength can be different and may even be reversed, counterintuitively.

\begin{figure}

\centering
\includegraphics[width=.5\textwidth]{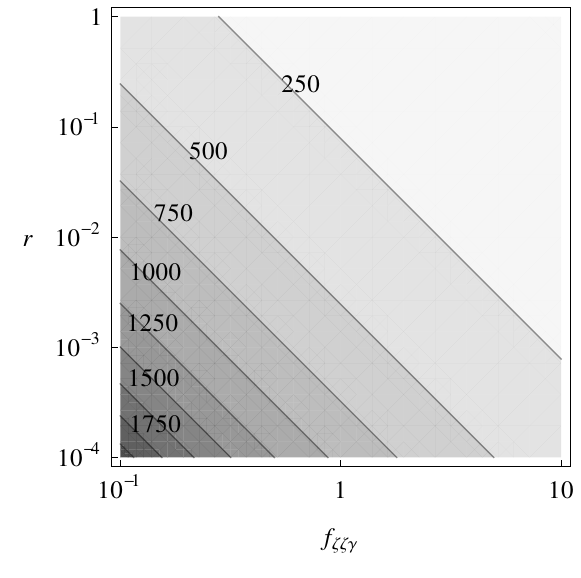}
\caption{Contour plot of $k_{\text{max}}/k_{\text{min}}$ for a $3\sigma$ detection of the gravitational fossil in off-diagonal contributions to the scalar power spectrum, in terms of the tensor-to-scalar ratio $r$ and scalar-scalar-tensor bispectrum amplitude $f_{\zeta\zeta\gamma}$, assuming a stronger-than-local squeezed limit ($m_S=1$, $m_L=-1$ in Eq. \eqref{squeezedinequality}). The signal is proportional to $f_{\zeta\zeta\gamma}^2 A_{\gamma}$. We set $r=A_{\gamma}/(2.2\times 10^{-9})$ and take $n_t\simeq0$.\label{PvsB}}

\end{figure}

In the case considered above of a primordial gravitational wave fossil field and non-Bunch-Davies initial state with constant Bogoliubov coefficients, $m_L=-1$ and $m_S=1$, leading to a stronger signal. Letting $f\rightarrow\gamma$ in Eq. \eqref{squeezedinequality} and setting $\alpha=3$ we can find the required $k_{\text{max}}/k_{\text{min}}$ for a $3\sigma$ detection. In ~\ref{PvsB} we show the dependence of this number on $f_{\zeta\zeta\gamma}$ and the tensor-to-scalar ratio.\footnote{We include the angular dependent factor $\epsilon^p_{ij}\hat{k}^i\hat{k}^j$ to the bispectrum with a spin-2 polarization tensor as described above in Section \ref{sec:estimator}, which leads to an extra factor of $\frac{4}{15}$ in Eq. \eqref{squeezedinequality}.} As expected, increasing either the bispectrum or tensor power spectrum increases the level of detectability of the tensor modes. For a sufficiently large three-point correlation with $K^{-4}$ behavior in the squeezed limit, and large enough tensor fluctuations, the signal would be within the observable range.

For the Bogoliubov initial state, we have $f_{\zeta\zeta\gamma}=\mathcal{O}(1)\times\sqrt{F^{-1}(\beta^{(s)},\beta^{(t)},\Theta^{(s)},\Theta^{(t)})}$. Thus, each point in Figure \eqref{PvsB} at $r=0.1$ or $r=0.01$ corresponds to a one-parameter family of values of $\beta^{(s),(t)}$ in Figure \ref{phases00}, if we fix $\Theta^{(s),(t)}$.

\section{Quadrupolar anisotropy and superhorizon modes}
\label{sec:quadrupole}

We have computed the fossil signature in off-diagonal correlations of the scalar power spectrum, which is insensitive to superhorizon tensor modes, since we cannot resolve scalar modes $\mathbf{k}_2$ and $\mathbf{k}_3$ separated by $|\mathbf{k}_2+\mathbf{k}_3|=K<a_0 H_0$. However, the tensor nature of the gravitational fossil can also give rise to a quadrupolar modulation of the scalar power spectrum \cite{Dai:2013kra}, introducing both statistical inhomogeneity and anisotropy. This signature is sensitive to superhorizon fossil modes if the squeezed-limit bispectrum is strong enough.

Consider a general scalar-scalar-tensor bispectrum with squeezed limit
\ba
\label{squeezed}
\lim_{k_L\ll k_S} B_p(\mathbf{k}_L,\mathbf{k}_S,-\mathbf{k}_L-\mathbf{k}_S)&&\nonumber\\
&&\hspace{-3cm}= f_{\zeta\zeta\gamma}P_{\zeta}(k_S)P^p_{\gamma}(k_L)\nonumber\\
&&\hspace{-2.8cm}\times\left(\frac{k_L}{k_P}\right)^{m_L}\left(\frac{k_S}{k_p}\right)^{m_S}\epsilon^p_{ij}\hat{k}_S^i\hat{k}_S^j,
\ea
where $k_p$ and $k_P$ are pivot scales, which we allow to be different as we are interested in $k_L$ and $k_S$ in different ranges,. ie. superhorizon $k_L$ coupling to subhorizon $k_S$. This bispectrum contributes to the scalar two-point function on a fixed tensor background, Eq. \eqref{offdiagonal}. Fourier transforming to real space, we can identify the power spectrum $P_{\zeta}(\mathbf{k})=\int d^3x' e^{-i\mathbf{k}\cdot\mathbf{x}'}\langle\zeta(\mathbf{x})\zeta(\mathbf{x}+\mathbf{x}')\rangle_{\gamma}$, as outlined in Appendix A of \cite{Dai:2013kra}. We find
\be\label{quadrupolepower}
P_{\zeta}(\mathbf{k},\mathbf{x})|_{\gamma}=P_{\zeta}(k)\left[1+f_{\zeta\zeta\gamma}\gamma_{ij}^{(m_L)}(\mathbf{x})\hat{k}^i \hat{k}^j\left(\frac{k}{k_p}\right)^{m_S}\right],
\ee
where
\be
\gamma_{ij}^{(m_L)}(\mathbf{x})=\sum_p \int_{L^{-1}} \frac{d^3K}{(2\pi)^3}e^{i\mathbf{K}\cdot\mathbf{x}}\gamma^p(\mathbf{K})\epsilon_{ij}^p\left(\frac{K}{k_P}\right)^{m_L}.
\ee
The quadrupolar anisotropy varies spatially with the tensor background, and depending on the squeezed limit behavior, it is more or less sensitive to longer-wavelength, superhorizon tensor modes. The infrared cutoff scale $L^{-1}$ is introduced because we do not expect Eq. \eqref{squeezed} to be valid for arbitrarily small $k_L$ (for example, bispectra computed from an effective field theory will be limited by the range of scales described by the theory). Note that the additional $k$-dependence from $m_S$ will affect the observed spectral index, an effect which was explored more generally in \cite{Bramante:2013moa}.

In the non-Bunch-Davies case with excited scalar modes $\beta_k^{(s)}\neq0$ overlapping with observable scales, $m_L=-1$ and the local quadrupole anisotropy is modulated by the longest wavelength ($\sim L$) tensor modes and thus varies less spatially. Note that whereas the off-diagonal fossil signature computed in Section \ref{sec:2} probes the correlation of observable scalar modes with longer but sub-Hubble tensor modes, whereas the quadrupole signature probes their correlation with superhorizon tensor modes. Both effects, however, originate in the excitation of the scalar modes in the initial state.

\section{Conclusion}
\label{sec:conclusion}
We have computed the gravitational fossil signature in off-diagonal correlations in the scalar two-point function, from a primordial scalar-scalar-tensor three-point function $\langle\gamma^p(\mathbf{k}_1)\zeta(\mathbf{k}_2)\zeta(\mathbf{k}_3)\rangle$ for excited Bogoliubov initial states for both the scalar and tensor modes. The bispectrum is largest in the squeezed and flattened limit $k_2\approx k_3\gg k_1$, $k_2+k_1\approx k_3$ (or $k_2-k_1\approx k_3$), characterized by an unusually strong $k_1^{-4}$ dependence, under the assumption that $\alpha^{(s),(t)}$ and $\beta^{(s),(t)}$ are constant. The fossil signature is obtained by summing over all measurable off-diagonal pairs of scalar modes. Depending on the amplitudes and phases of the Bogoliubov coefficients, the survey size needed for a detection can be as small as $k_{\text{max}}/k_{\text{min}}\sim10^2$, within the reach of current surveys.

We have also given, in Section \ref{sec:fossilsqueezedlimit} a phenomenological parameterization for the off-diagonal fossil signature in terms of the amplitude of fossil fluctuations, and amplitude and squeezed-limit scaling $\frac{d\ln B}{d\ln k_L}$ of a scale-invariant scalar-scalar-fossil bispectrum. This illustrates the necessary criteria for any primordial correlation to satisfy in order to leave an observable imprint. The mode coupling introduced by the initial state can also lead to a gravitational fossil signature in the form of a quadrupole anisotropy in the scalar power spectrum, with greater sensitivity to longer wavelength tensor modes, as described in Section \ref{sec:quadrupole}.

Here we have studied the effect on the scalar and tensor perturbations from a non-Bunch-Davies initial state imprinted by unknown pre-inflationary dynamics. In the Bunch-Davies case the clock of the inflationary background dynamics determines the statistics of the fluctuations. The attractor behavior forbids correlations forming with modes that have crossed the horizon, and correlations forming with very subhorizon modes are exponentially suppressed \cite{Flauger:2013hra}. Consequently, correlations between modes of very different wavelength are disallowed, and the squeezed-limit bispectrum vanishes. Pre-inflationary dynamics, on the other hand, can result in excited modes at the onset of inflation so that the amplitude of fluctuations on scale $k$ is not purely determined by the background evolution ($H^2/\sqrt{\epsilon}$ when the mode crosses out of the inflationary horizon). If subhorizon scalar modes are excited, they can now correlate to longer tensor modes when the tensor modes cross the horizon, before they become part of the classical background, resulting in a large squeezed-limit bispectrum. From Eq. \eqref{bispectrum} we see that $\beta_k^{(s)}\neq0$ is the necessary and sufficient condition for the $B\propto k^{-2}K^{-4}$ squeezed limit scaling ($\beta^{(t)}\neq0$ allows the tensor modes to be influenced by an additional clock, and to correlate with other modes before crossing the horizon, but they could not couple to shorter scalar modes unless they too were excited). In the same way, in order for short tensor modes to couple to long scalar modes through the $\zeta\gamma\gamma$ interaction, the tensor modes would have to be excited when deep inside the horizon, requiring that $\beta_k^{(t)}\neq0$. (Classically, a realization of short-wavelength modes of one field $X_S$ can be adjusted to be correlated with a fixed realization of long-wavelength modes of a field $X_L$, but not vice versa. So it is the $X_S$ modes that must be influenced by dynamics other than those of the inflationary background if they are to couple to $X_L$.)

A new clock influencing the fluctuations can also come from additional dynamics \textit{during} inflation from a non-inflaton sector. For example, although tensor perturbations always arise from the vacuum, they can also be generated via particle production and decay of other fields during inflation. Depending on whether this production dominates over the vacuum fluctuations and on the relation between the inflaton dynamics and dynamics governing particle production, the scalar and/or tensor perturbations may be governed by a clock other than the inflationary background. The usual intuition that non-Gaussianity in the gravitational sector is small compared to that in the scalar sector doesn't have to hold, as was recently shown by \cite{Cook:2013xea}. In the typical particle production scenario the non-Gaussianity is not of the local type, but it would be very interesting to explore the range of physical mechanisms, and conditions on additional dynamics before or during inflation that would be necessary to generate squeezed-limit mode coupling of scalar and/or tensor modes. For example, one might ask if governing both scalar and tensor perturbations with the same dynamics, but different from the inevitable fluctuations from inflation, could remove any coupling between scalar and tensor modes. If this is possible, the tensor modes may not be as clear a diagnostic of the inflationary background (and classical evolution of the inflaton) as we currently hope they are.

The fossil signature studied here comes from the minimal gravitational coupling of the inflaton. Fossil signatures may also come from coupling of the inflaton to other sectors or from nonminimal coupling to gravity. It could be interesting to investigate conditions on such couplings that allow for an observable fossil signature, subject to observational constraints, for example on isocurvature modes.

Modulation of local statistics by a fossil field may also be worth investigating in light of anisotropy features on large scales in the CMB \cite{Ade:2013nlj}. Superhorizon scalar curvature modes coupled to scalar modes on observable scales with an anisotropic $\langle\zeta\zeta\zeta\rangle$ bispectrum can lead to an observable power asymmetry \cite{Schmidt:2012ky}. It would be interesting to see under what conditions the squeezed-limit coupling to superhorizon modes of another field can be strong enough to give rise to such effects. A non-Bunch-Davies initial state or additional dynamics before or during inflation may introduce greater sensitivity of $\zeta$ to background modes of other fields, as seen here for tensor modes.
\\
\section{Acknowledgments}

We thank Nishant Agarwal, Liang Dai, Donghui Jeong, Marc Kamionkowski, and Sonia Paban for helpful correspondence and discussions. The work of E.N. and S.S. is supported by the New Frontiers in Astronomy and Cosmology program at the John Templeton Foundation. S.B. is supported by NSF grant PHY-1307408.

\bibliographystyle{JHEP}

\bibliography{Fossil}

\end{document}